\documentclass{article}
\usepackage{dcase2024_techrep,amsmath,graphicx,url,times,booktabs, tabularx}
\usepackage{latexsym}
\usepackage[T1]{fontenc}
\usepackage[utf8]{inputenc}
\usepackage{microtype}
\usepackage{inconsolata}
\usepackage{amsfonts}
\usepackage{multirow}
\usepackage{arydshln}
\usepackage{xspace}

\newcommand{\etal}{\textit{et~al.}\xspace}

\title{Expanding on EnCLAP with Auxiliary Retrieval Model for Automated Audio Captioning}

\name{Jaeyeon Kim$^{1,2}$,
      Jaeyoon Jung$^{2,3}$,
      Minjeong Jeon$^{2}$, 
      Sang Hoon Woo$^{4}$,
      Jinjoo Lee$^{2}$, 
      }
      
\address{$^1$ Seoul National Unversity, Seoul, Republic of Korea \\
        $^2$ MAUM AI Inc., Seongnam, Republic of Korea, \\
        $^3$ Soongsil University, Seoul, Republic of Korea, \\
        $^4$ Independent Researcher \\
        \texttt{jaeyeonkim99@snu.ac.kr \{jyjung, mjjeon, jjl\}@maum.ai tonyswoo@gmail.com}
 }

\begin{document}

\ninept
\maketitle

\begin{sloppy}

\begin{abstract}
In this technical report, we describe our submission to DCASE2024 Challenge Task6 (Automated Audio Captioning) and Task8 (Language-based Audio Retrieval). We develop our approach building upon the EnCLAP audio captioning framework and optimizing it for Task6 of the challenge. Notably, we outline the changes in the underlying components and the incorporation of the reranking process. Additionally, we submit a supplementary retriever model, a byproduct of our modified framework, to Task8.
Our proposed systems achieve FENSE score of 0.542 on Task6 and mAP@10 score of 0.386 on Task8, significantly outperforming the baseline models.
\end{abstract}


\section{Introduction}
\label{sec:intro}

Automated audio captioning (AAC) refers to the cross-modal translation task of transcribing audio signals that contain sound events into concise and meaningful natural language descriptions \cite{aac}. 
Despite the recent success of deep learning in many traditional tasks, AAC remains a particularly challenging task, with substantial performance discrepancy between human and machine. 

One significant contributor of the performance gap can be attributed to the intrinsic complexity of the task. 
Distinguishing between various sound events, especially between similar and ambiguous ones, requires extensive real-world knowledge. 
To mitigate this challenge, prior studies have incorporated additional real-world acoustic knowledge by employing pretrained audio encoders trained on audio classification tasks \cite{mei, conette, beats-conformer}. 

The scarcity of high-quality data poses an additional challenge in audio captioning.
Notably, AudioCaps \cite{audiocaps} and Clotho \cite{clotho}, two most widely used datasets for audio captioning, contain approximately 50K and 20K captions, respectively, while COCO captions \cite{coco_captions}, a widely used dataset for image captioning, has over 414K captions in its training set. 
Although Mei \etal \cite{wavcaps} proposed WavCaps, a large-scale audio captioning dataset comparable in scale to COCO captions, it is important to note that WavCaps is a weakly-labeled dataset and cannot be considered a direct substitute for a high-quality dataset. 
To address this issue, previous works \cite{prefix_tuning, pengi, gontier} have leveraged the text generation capabilities of pretrained language models like GPT-2 \cite{gpt2} and BART \cite{bart} to improve the semantic quality of the captions under data-scare scenarios. 
Additionally, some studies have also incorporated auxiliary loss terms, including keyword prediction loss \cite{koizumi_keyword} or sentence embedding loss \cite{sentence_embedding}, to provide additional training signal and improve the training procedure.

Expanding upon previous line of works, Kim \etal \cite{enclap} proposed the EnCLAP framework, which integrates a set of pretrained models with an auxiliary training task. 
Notably, EnCLAP utilizes two acoustic feature encoders, EnCodec \cite{encodec} and CLAP \cite{clap_laion}, to generate timestep-level and sequence-level representation, of the input audio sequence, respectively. 
For caption decoder, the framework employs a pretrained BART model \cite{bart}. 
Furthermore, Kim \etal also introduced masked codec modeling (MCM), an auxiliary task designed to enhance the acoustic awareness of the caption decoder. 
The combination of these approaches allowed EnCLAP to achieve state-of-the-art performance on the AudioCaps dataset.

In this work, we adapt the EnCLAP framework to tackle DCASE2024 Challenge. 
We aim to optimize and enhance the performance of each component within the EnCLAP framework, while adhering to the challenge's rules and regulations. 
Specifically, we investigate alternative models for the EnCodec and CLAP components and adopt a sampling and reranking procedure to further improve the quality of the generated captions. 
We submit our resulting system to Task6 and Task8 of the challenge. 

\begin{table*}[]
    \label{table:retrieval}
    \caption{Results on Task8 Language-Based Audio Retrieval on Clotho evaluation split. CL, AC, and WC refer to Clotho, AudioCaps, and WavCaps, respectively.}
    \vspace{0.5\baselineskip}
    \centering
    {
    \begin{tabular}{ccccccc}
    \hline
    Model & Audio Encoder & Text Encoder & mAP@10 & R@1 & R@5 & R@10 \\
    \hline
    Baseline & CNN14 & all-mpnet-base-v2 & 0.222 & 0.130 & 0.343 & 0.480 \\
    \cdashline{0-6} 
    \multicolumn{7}{c}{\textit{Pretrained on CL + AC + WC}} \\
    Pretrain 1 & CNext & bge-base & 0.334 & 0.222 & 0.485 & 0.619 \\
    Pretrain 2 & CNext & bert-base & 0.325 & 0.208 & 0.479 & 0.618 \\
    Pretrain 3 & CNext & roberta-base & 0.326 & 0.214 & 0.474 & 0.614 \\
    Pretrain 4 & CNext & bge-large & 0.339 & 0.219 & \textbf{0.502} & 0.633 \\
    Pretrain 5 & CNext & bert-large & 0.339 & 0.220 & 0.500 & \textbf{0.635} \\
    Pretrain 6 & CNext & roberta-large & \textbf{0.342} & \textbf{0.228} & 0.492 & 0.632 \\

    \cdashline{0-6}
    \multicolumn{7}{c}{\textit{Finetuned on CL}} \\
    Finetune 1 & CNext & bge-base & 0.356 & 0.235 & 0.522 & 0.649 \\
    Finetune 2 & CNext & bert-base & 0.356 & 0.235 & 0.520 & 0.654 \\
    Finetune 3 & CNext & roberta-base & 0.365 & 0.250 & 0.523 & 0.659 \\
    Finetune 4 & CNext & bge-large & 0.369 & 0.249 & 0.530 & 0.665 \\
    Finetune 5 & CNext & bert-large & 0.367 & 0.247 & 0.526 & 0.663 \\
    Finetune 6 & CNext & roberta-large & 0.375 & 0.256 & 0.535 & 0.669 \\
    \cdashline{0-6}
    Ensemble 1 & & & 0.385 & 0.265 & \textbf{0.547} & 0.676 \\
    Ensemble 2 & & & \textbf{0.386} & \textbf{0.267} & \textbf{0.547} & \textbf{0.680} \\
    Ensemble 3 & & & 0.378 & 0.257 & 0.543 & 0.676 \\
    \hline
    \end{tabular}}
    \label{tab:my_label}
\end{table*}
\section{Method}
\label{sec:method}

\subsection{Neural Audio Codec}
\label{subsec:encodec}

Neural audio codecs are autoencoder models designed to encode waveforms into sequences of discrete codes. 
Recent advancements \cite{soundstream, encodec, dac} typically employ residual vector quantization (RVQ) for compression, utilizing multiple codebooks to quantize the residuals of preceding codebooks. 
Ultimately, the input waveforms are transformed into a set of parallel discrete code sequences, each of which is associated with a unique codebook.
Neural audio codecs have demonstrated success as the acoustic representation format in generative audio models \cite{kreuk2022audiogen, valle, musicgen}. 

In the context of audio captioning, EnCLAP \cite{enclap} employs the neural audio codec, specifically EnCodec \cite{encodec}, to represent the input waveform at the timestep level. 
This approach is based on the assumption that pretrained language models are better suited to process discrete inputs compared to continuous ones. 
In this work, we replace EnCodec in the original EnCLAP framework with Descript Audio Codec (DAC) \cite{dac}, as DAC has demonstrated superior performance in audio compression, as well as downstream tasks \cite{dac, codec_superb}. 

\subsection{Audio-Text Joint Embedding}
\label{subsec:clap}
The original EnCLAP employs CLAP \cite{clap_laion} embeddings as the sequence-level acoustic representation of the input audio. 
However, due to potential overlap between the training dataset of CLAP and the evaluation dataset, we substitute CLAP with an alternative model.
Specifically, we utilize the audio encoder of the baseline model of the challenge, hereinafter referred to as CNext \cite{convnext}, which was trained on the AudioSet \cite{audioset} dataset for the audio classification task. 
In our preliminary experiments, we observed that the variant of CNext finetuned using the audio-text retrieval task exhibits superior performance. 
Therefore, we adopt this variant in our work. 

\subsection{Generation and Reranking}

Previous works, including EnCLAP \cite{enclap}, have utilized beam search decoding for caption generation. 
However, Wu \etal \cite{beats-conformer} showed that the sampling-then-reranking approach yields more diverse and informative captions. 
Therefore, we adopt the approach proposed by Wu \etal \cite{beats-conformer}, where we generate a set of candidate captions through nucleus sampling and select the most suitable one via reranking. 

Our candidate selection procedure is a two-stage process. 
First, we use the FENSE fluency error detector \cite{fense} to filter captions containing fluency errors. 
We then rank the remaining candidates based on the weighted sum of two reranking scores: encoder reranking and decoder reranking. 
The encoder reranking score is cosine similarity score between the input audio representation and the generated caption representation computed using the retriever model described in Section \ref{subsec:clap}. 
For the decoder reranking score, we use the log-likelihood of the generated caption given the input audio.

\begin{table*}{
\centering
\caption{Results on Task6 Automated Audio Captioning on Clotho evaluation split. For EnCLAP-large, we report the scores using the official \textit{clotho-finetune-large} checkpoint, which was pretrained on the AudioCaps dataset and finetuned on the Clotho dataset.}
\label{table:aac}
\begin{center}
{
\begin{tabular}{cccccccc}
\hline
Model & METEOR & CIDEr & SPICE & SPIDEr & SPIDEr-FL & Vocabulary & FENSE \\
\hline
Baseline & 0.1897 & 0.4619 & 0.1335 & 0.2977  & 0.2962  &551  & 0.5040 \\
EnCLAP-large & 0.1864 & 0.4641 & 0.1336  & 0.2989 & 0.2971 & 592 & 0.5116 \\
\cdashline{0-7}

Submission1 & 0.1989 & 0.\textbf{4826} & 0.1483 & \textbf{0.3155} & \textbf{0.3155} & 840 & 0.5386 \\
Submission2 & 0.1955 & 0.4775  & 0.1423 & 0.3099 & 0.3099 & \textbf{865} & 0.5419 \\
Submission3 & \textbf{0.2003} & 0.4780 & \textbf{0.1488} & 0.3134 & 0.3134 & 825 & 0.5393 \\
Submission4 & 0.1994 & 0.4778 & \textbf{0.1488} & 0.3133 & 0.3133 & 815 & \textbf{0.5420} \\
\hline
\end{tabular}
}
\end{center}
}\end{table*}
\section{Experiment}
\label{sec:experiment}
We assess the performance of our modified EnCLAP model on Task6 of DCASE2024 Challenge and report the results. Additionally, we evaluate the retriever model described in Section \ref{subsec:clap} on Task8 of the challenge. 

\subsection{Setup}
\label{subsec:setup}

\textbf{Dataset.} In our experiment, we adopt a two-stage training process, where we pretrain on a larger dataset and subsequently finetune on a smaller dataset. 
The pretraining dataset comprises a combination of AudioCaps \cite{audiocaps}, WavCaps \cite{wavcaps}, Clotho \cite{clotho}, and Clotho-ChatGPT-Mixup \cite{beats-conformer}. 
Conversely, the finetuning dataset consists solely of the Clotho dataset. To comply with the challenge regulations, we exclude any potential overlapping data from Freesound in the WavCaps dataset. Additionally, we only use audio clips with durations between 1 and 30 seconds from the WavCaps dataset. For Clotho, we utilize only the training split of the dataset. 

\noindent\textbf{Model Configuration.} From the original EnCLAP model configuration, we experiment only with the EnCLAP-large setup to maximize the performance of the final model. We use a variant of the DAC model that transforms a 24kHz acoustic waveform to 75Hz code sequences, with 32 codes per timeframe and codebook size of 1024. For the retriever model, we initialize the audio encoder with CNext-tiny by Pelligrini \etal \cite{convnext}. For the text encoder, we experiment with 6 different pretrained language models: BGE-base \cite{bge}, BERT-base \cite{bert}, RoBERTa-base \cite{roberta}, BGE-large \cite{bge}, BERT-large \cite{bert}, RoBERTa-large \cite{roberta}. For the sequence-level feature encoder, we choose Pretrain 1 version listed in Table 1. Note that we resample the input audio to appropriate sample rates before processing it with feature encoders.

\noindent\textbf{Generation.} We use nucleus sampling with a probability threshold of 0.95 and a temperature of 0.5 to generate 30 candidates. 
We rank the candidates by the weighted sum of the encoder reranking score and the decoder reranking score using weights of 0.7 and 0.3, respectively.

\subsection{Training}
 
\textbf{Language-Based Audio Retrieval.} We train our retriever models using the m-LTM framework \cite{m-ltm}, a learning-to-match framework for the minibatch setting, designed to minimize the modality gap between audio and text embedding in audio-text retrieval tasks. 
During the pretraining phase, we use a mixed dataset comprising AudioCaps \cite{audiocaps}, WavCaps \cite{wavcaps}, and Clotho \cite{clotho}.

\noindent\textbf{Automated Audio Captioning.} For audio caption training, we follow the original EnCLAP setup and use a combination of two tasks: captioning task and MCM task. MCM is an auxiliary training task which involves masking a part of the input codec sequence and predicting it, analogous to the masked language modeling (MLM) approach.
Note that we omit the MCM task during the finetuning stage. 
During the pretraining stage, we use a combined dataset of AudioCaps \cite{audiocaps}, WavCaps \cite{wavcaps}, and Clotho-ChatGPT-Mixup \cite{beats-conformer}.

\subsection{Results}
\label{subsec:results}
\textbf{Language-based Audio Retrieval. } We present the results of our evaluation on Task8 in Table \ref{table:retrieval}. 
Our models significantly outperform the baseline.
We also find that ensembling models with different encoders yields additional score improvements. 
For the challenge, we submit the following four models:
\begin{enumerate}
    \item \textbf{Finetune 1:} CNext audio encoder and RoBERTa-large text encoder
    \item \textbf{Ensemble 1:} An ensemble of the top 3 fine-tuned models: Finetune 4, 5, 6
    \item \textbf{Ensemble 2:} An ensemble of all fine-tuned models
    \item \textbf{Ensemble 3:} An ensemble of all pre-trained and fine-tuned models
\end{enumerate}

\noindent\textbf{Automated Audio Captioning. }We summarize the results of our evaluation on Task6 in Table \ref{table:aac}. 
Our models surpass both the DCASE2024 baseline and EnCLAP-large by a wide margin.
The details of our submissions are as follows: 
\begin{enumerate}

    \item \textbf{Submission 1:} A modified EnCLAP model with DAC and CNext audio-text joint embedding 
    \item \textbf{Submission 2:} An average soup \cite{modelsoup} model of 5 modified EnCLAP models
    \item \textbf{Submission 3:} An ensemble of 7 modified EnCLAP models
    \item \textbf{Submission 4:} An ensemble of 7 modified EnCLAP models and 2 soup models
\end{enumerate}

\section{Conclusion}
\label{sec:conclusion}
This report outlines our approach to DCASE2024 Challenge. We investigate the application of the recently introduced m-LTM loss to language-based text retrieval. For automated audio captioning,  we attempt to optimize and improve the the EnCLAP framework by introducing new backbone models, that is, specifically by replacing EnCodec with DAC and CLAP with the audio encoder from the aforementioned retriever. We also integrate a sampling-and-reranking scheme to the generation procedure. In our future work, we hope to investigate the reciprocal effects of captioning and retrieval tasks.

\bibliographystyle{IEEEtran}
\bibliography{refs}

\end{sloppy}
\end{document}